\documentclass[final,5p,times,twocolumn]{elsarticle}
\pdfoutput=1 
\usepackage{graphicx}

\usepackage{amssymb}

 \usepackage{lineno}

\journal{Nuclear Instruments and Methods A}

\begin{document}

\begin{frontmatter}



\title{An assembly line for the construction of ultra-radio-pure detectors}


\author [1]{E. Buccheri}
\author [1]{M. Capodiferro}
\author [1]{S. Morganti}
\author [1]{F.Orio}
\author [1]{A. Pelosi}
\author [1,2]{V. Pettinacci}

\address[1]{Istituto Nazionale di Fisica Nucleare, Sezione di Roma, P.le A. Moro 2, Roma, Italy}
\address[2]{Department of Physics and Astronomy, University of South Carolina, Columbia, USA}

\begin{abstract}
The 19 towers constituting the CUORE detector are an assembly of ultra-radio-pure components made of copper and PTFE plus tellurium dioxide crystals for a total of more than 10,000 pieces. A dedicated assembly line is mandatory to handle and assemble those parts and to minimize the risk of recontamination induced by external agents both during their construction and during their storage prior to installation inside the cryostat. The assembly strategy and the tool design proposed in this paper and developed for CUORE offer solutions that can be extended to the construction of similar-size detectors with strict requirements regarding contamination.
\end{abstract}

\end{frontmatter}

\section{Introduction}

\begin{figure}[h]
\begin{centering}
\includegraphics[width=\columnwidth]{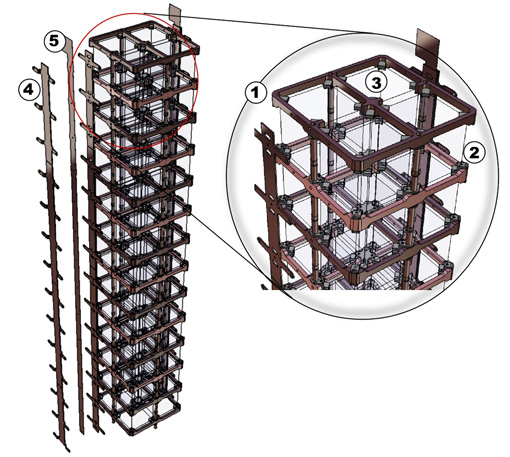}
\caption{The CAD model shows the main components the tower modular structure is made of: copper frames (1); corner, central and side PTFE holders (2); crystals (3); wire trays (4); and the PCB strips carrying the wiring.}
\label{fig:figura_1}
\end{centering}
\end{figure}

CUORE (Cryogenic Underground Observatory for Rare Events) is an upcoming cryogenic experiment that is assembled and to be installed at the LNGS (Laboratori Nazionali del Gran Sasso) underground facility to observe neutrino-less double beta decay ($\beta\beta0\nu$) ~\cite{1}, \cite{2}, \cite{3}. It consists of 988 ultra-pure \cite{4} TeO$_{2}$ cubic crystals (50x50x50~mm) that are operated as bolometers and grouped into 19 towers with 52 crystals each. Crystals in a tower are held in groups of four by 14 thin radioactive-pure copper frames spaced by columns made of the same material. A set of small PTFE holders (see figure ~\ref{fig:figura_1}) 
keep the crystal in place and maintain the mechanical stress applied to them at the cryogenic temperature. The 19 towers, which are assembled to form a very compact structure and operated at 10~mK inside an ad hoc designed cryostat, will constitute the CUORE detector. A complete description of the CUORE experiment can be found in \cite{5}.\\
The dominant sources of background in CUORE are the surface and bulk radioactive contaminations of constituting materials. These are due to natural impurities or contaminants of external origin that may adhere to or become embedded in the exposed surfaces of the material. To reach the CUORE target background of 0.01~counts/keV/kg/y in the $\beta\beta0\nu$ region of interest (ROI), all of these sources must be very carefully addressed along the whole construction by considering the following points:
\begin{itemize}
\item An accurate selection of raw materials (origin, composition, production technique); 
\item The minimization of their permanence aboveground (for example, for parts machining and cleaning) to avoid cosmic ray activation;
\item The use of clean protocols and tools to machine and shape the components;
\item The recognition of the most effective final cleaning process to strike down residual surface contamination ~\cite{6};
\item The design of an assembly line providing all of the tools and procedures required to preserve the extremely high level of radio-purity of the final components.
\end{itemize}
\section{CUORE towers assembly line outline}

The inputs to the CUORE towers assembly line (CTAL) are approximately 10.000 ultra-clean pieces, which are produced, machined and cleaned according to the aforementioned recommendations. The CTAL task is to transform those parts into 19 ultra-clean towers ready to be installed inside the cryostat.\\
The main requirements are as follows:
\begin{itemize}
\item Contain to a minimum the recontamination risk due to direct contact with contaminated (or simply less radio pure) materials and exposure to solid (any powder) or gaseous (radon) contaminants in the atmosphere;
\item Design a system to guarantee the protection of the tower and its cleanness even for a long period; 
\item Ensure that the assembly procedure is reversible;
\item Minimize the assembly time;
\item Design the system to fit inside the CUORE clean room.
\end{itemize}

\begin{figure}[!]
\begin{centering}
\includegraphics[width=\columnwidth]{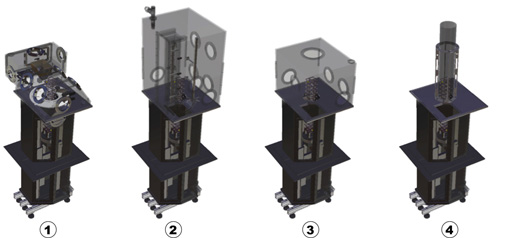}
\caption{The four glove boxes that can be mounted on the universal working plane coupled with the "garage". }
\label{fig:figura_2}
\end{centering}
\end{figure}

\begin{figure}[!]
\begin{centering}
\includegraphics[scale=1]{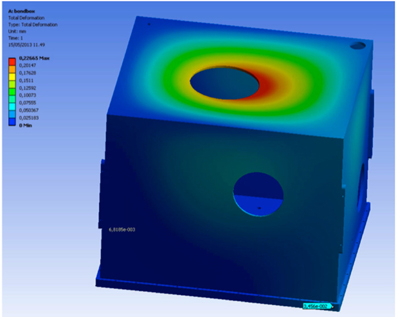}
\caption{Ansys calculated the deformation and stress of the bonding box induced by the ~40 kg load of the bonding machine. A limited deformation (0.2 mm) is expected excluding shape variations dangerous for the O-Ring (1 mm squeezed in standard condition), and a compatible stress distribution on the PMMA body with a maximum principal stress value of approximately 5 MPa. }
\label{fig:figura_3}
\end{centering}
\end{figure}

In addition, the design of the CTAL has to take into account the fact that the tower reaches its final structural rigidity only at cryogenic temperatures. It is actually at these temperatures that the tower elements have been dimensioned on the basis of the different thermal contractions of the materials they are made of. A very smooth and careful handling of the structure is then required to avoid mechanical shock or acceleration that could damage the tower.\\
To cope with the minimum recontamination requirement, CTAL has been conceived as a "zero contact" assembly line. The tower cleaned components are protected at any stage of the assembly process from direct contact with objects made of non-radio pure materials and with solid or gaseous (radon) contaminants in the atmosphere. Their handling is minimized and is always performed by tools made of certified low activity materials.\\
To work in a clean, low-radon, N$_{2}$  atmosphere, the whole assembly process is confined into hermetic volumes accessible only through sealed glove ports. The working volume is split into two intercommunicating, independently sealed zones. One zone is fixed, shared by all production tools, and the other is interchangeable to face different assembly operations. The fixed volume is a lift/storage subsystem called a "garage". It is used to host the tower under construction throughout the assembly period and to provide two motion axes (rotation and vertical translation) to the assembly plate. The UWP, a "universal" working plane (i.e., common to all of the assembly tools and stages), fixed to the garage top flange provide the mechanical coupling interface to the second volume, which can be one of four different glove boxes 
(see figure~\ref{fig:figura_2}).
The garage and the glove-box mounted on the UWP are independently sealed and fluxed with evaporated nitrogen. The two volumes communicate through an aperture on the UWP, allowing the transfer of the detector under construction from its storage location (the garage) to the glove-box and vice versa.\\
Mounted on the UWP, a specific glove-box performs any of the following assembly stages:
\begin{itemize}
\item Mechanical Assembly: copper frames and crystals are assembled in a meck box (n.1 in figure 2), forming the main structure.
\item Cabling: the copper wire trays, supporting a flexible printed circuit (PCB) used to drive out bolometer signals, are installed on 2 opposite faces of the tower using a cabling box (n.2 in figure 2).
\item Bonding: the bonding box (n.3 in figure 2) hosts a modified Westbond wire bonder operated to connect the sensors glued on each crystal to the cabling strips.
\item Covering: making use of the cabling box, copper covers are placed over the wire trays to shield and secure the PCB strips.
\item Storage: the finalized tower is stored inside a sealed canister (storage box, n.4 in figure 2). The 19 nitrogen-fluxed boxes are parked in a rack to await installation inside the cryostat.
\end{itemize}
The four interchangeable glove boxes, which are designed considering the ergonomics and the set of tools and instruments required by the specific task, have completely different form factors and operator access points.\\
The assembly line is completed by a fifth stand-alone glove box with a much larger working space and multi operator access ports. This special glove box (End Of the World Box, EOW-Box in the following) is dedicated to all pre-assembly operations (extraction of the components from their storage packaging, parts quality check, run-in, pre-assembly of tower components subsets, etc., as well as their reverse). Its design and associated tools provide a very flexible system capable of facing unexpected or non-standard operation. 
All of the glove boxes are made of PMMA, which offers an optimal compromise between transparency, low radon permeability \cite{7}, density and cost. Finite element analysis has been used to find the optimal compromise between the strength and the weight of each glove box, while the Rankine-Navier criterion has been adopted to verify the stress distribution with a safety factor of SF = 5 to prevent any risk of structure failure.\\
The need to place the ~40 kg wire bonder on the UWP overhang portion required a complete static-structural analysis 
(figure ~\ref{fig:figura_3})
to calculate the deformation and stress induced on the UWP and on the glove box's body and confirm that the sealing between the two is not affected.
\section{CTAL installation in the LNGS Clean Rooms }
\begin{figure}[h]
\begin{centering}
\includegraphics{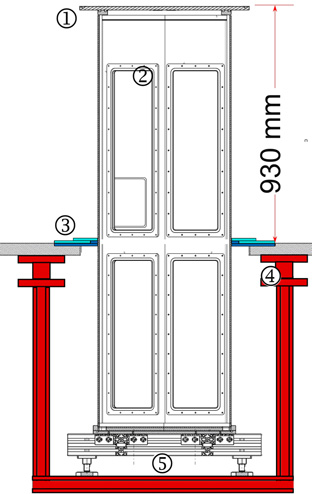}
\caption{Side view of the garage installation in clean room. 1) The UWP at 930 mm from the floor; 2) the garage body; 3) the three floor flanges sealing the breach in the floor; 4) the cage hosting the garage bottom; 5) the garage stand.  }
\label{fig:figure_4}
\end{centering}
\end{figure}

The assembly line is installed and operated in a clean room on the first floor of the CUORE building at the LNGS underground facility and is named "HUT". The single assembly working station with interchangeable glove boxes works well inside the limited space allocated for the assembly line. The total height of the garage structure is determined by the needs to host a complete tower (934 mm) plus its mechanical coupling to the lift/rotation table and its motion actuators (figure ~\ref{fig:figure_4}).
To ensure the UWP is at the correct height, the garage stand on the cage is fixed below the building's main structure, and the clean room protrudes through a square aperture on the floor.
A three-flange system seals the breach and restores clean room integrity.
\section{The Universal Working Plate}
\begin{figure}[h]
\begin{centering}
\includegraphics{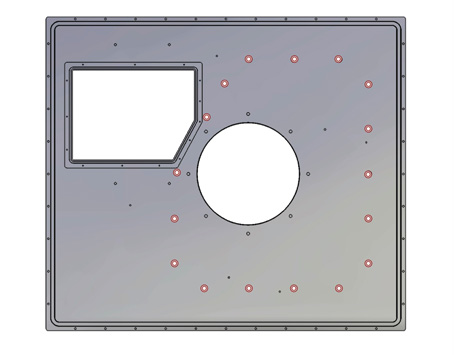}
\caption{The UWP tools interface. In red are shown the connections with the garage shell and in black the hole patterns shared by the different tool sets used at different stages of the assembly. }
\label{fig:figura_5}
\end{centering}
\end{figure}

All of the assembly actions occur on the UWP within the N$_{2}$ atmosphere volume defined by the glove box mounted on top. Its surface (700x800 mm$^{2}$) determines the effective working area and is the result of optimization among ergonomics issues, operative constraints (two facing operators must be able to exchange an object), internal tools occupancy, and minimization of the system reconditioning time. The plane is obtained by machining a 15-mm-thick ALCOA MIC6 aluminum slab requiring a surface flatness tolerance of 0.38 mm. A silicon-rubber O-ring running all along the UWP border provides the sealing with the glove box.\\
Its rectangular shape has two apertures (see figure ~\ref{fig:figura_5})
 : a central hole open to the garage and a shaped window on the bottom-right corner to host the bonding machine. Its asymmetric profile maximizes the wire bonder's range of motion (see section 9), making all of the bonding targets on the tower easily accessible. An independent threaded holes pattern around the shaped window fastens the PMMA cover underneath, which completes the sealing of the UWP. If not used to host the wire bonder, the PMMA box underneath is used as a repository to free the working space from bulky objects such as the UWP's central lid.\\
The surface presents some patterns of the blind threaded holes used to fix the many different tools that, stage by stage, are needed for the assembly. Because those holes make it more difficult to properly clean the UWP surface, their number and positions have been carefully optimized. 
\section{The garage}
\begin{figure}[h]
\begin{centering}
\includegraphics{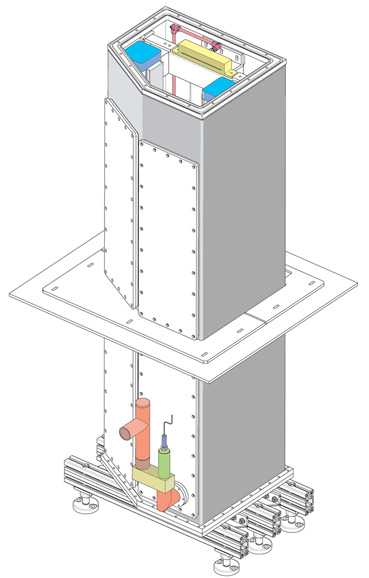}
\caption{The garage equipment. Two parallel FESTO linear actuators are fixed to a BOSH Retrox aluminum stiff structure providing the coupling points with the external shell. The top and bottom inductive position sensors cabled with the remote set the upper and lower limit of the travel a few millimeters before a mechanical security stop switch connected to the master security circuit. The rotation stage and its step motor are set on the lifting platform. The up-down, cw-ccw movements are controlled by a remote wired to the master electronic box located in the clean room. }
\label{fig:figura_6}
\end{centering}
\end{figure}

\begin{figure}[h]
\begin{centering}
\includegraphics{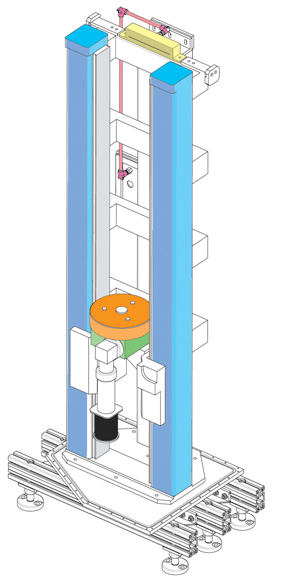}
\caption{The garage shell. The relief valve and the polarimetric sensors at the gas outlet are on the bottom. A flange on the rear, at the clean room level, hosts the pass-through plugs for the power, step motor driving pulses and safety sensors plus the gas inlet plug.}
\label{fig:figura_7}
\end{centering}
\end{figure}

The garage is the key CTAL volume. It is the N$_{2}$ fluxed main storage for the tower under construction, both while swapping the glove box from an assembly stage to the next and in the case of long term production stops. It is equipped with a motorized linear actuator (assembly vertical motion) with a rotation stage and the assembly plate on top (figure ~\ref{fig:figura_6}).
\\The two axis motions are driven by two independent SANYO step motors, one connected to two coupled parallel 1100 mm FESTO linear actuators and the other connected to the rotation stage. The assembly plate on the rotating plate matches a simple three-dowel system transmitting the spin shear force.
Two RTA MIND T series step motor drivers, generating a pulse train ranging from 3 kHz up to 24 kHz, control the step motors. The pulse frequency range corresponds, through proper reduction, to a maximum vertical speed of 38.4 mm/s and a minimum speed of 150 $\mu$/s, the latter being required for a very precise setting of the tower when aiming at the bonding target.\\
The 1.6 $\mu$m/step precision in the tower positioning has been exploited, in combination with an optical microscope system, to use the vertical motion to make a very precise and touch-less measurement of the assembly geometry, as described in section 7.\\
Two speed settings have also been implemented for the rotation: the lowest to let the bonding operator fine tune the angle between the wire bonder tip and the tower and the faster to rotate the assembly according to the operator's needs.\\
Detector construction starts on the assembly plate coupled with the rotation stage. A threaded M4 holes pattern is used to fix the assembly bottom or, as in the CUORE case, to mount a stand to raise the detector to the most convenient height for the next assembly stages. The maximum height allowed for the final assembly (detector plus stand) is 1100 mm.\\
The garage outer shell (figure ~\ref{fig:figura_7})
is made of 5-mm-thick AISI 304 sheet metal, which is bent to form its complex profile and TIG welded to make sure it is gas tight. Compared to methacrylate, stainless steel has a lower radon permeability, thus giving the assembly safer long-term storage within the CTAL chain. Four frontal PMMA windows allow for the visual inspection of the inside and access for maintenance work. The trapezoidal footprint follows the shape of the UWP square opening to host the UWP below the cover housing the wire bonder bottom.\\
On the garage top flange, a double concentric O-ring seals the garage body to the UWP.\\
The garage is continuously purged with evaporated nitrogen at a rate of 800 l/h from a diffuser on the top, and its internal environment purity is constantly monitored by a polarimetric oxygen sensor placed at the gas outlet pipe on the bottom side, which is outside the clean room.
\section{Glove box design and common subsystems }
\begin{figure}[t]
\begin{centering}
\includegraphics{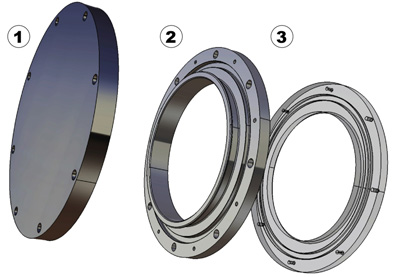}
\caption{The glove-port. The PMMA flange (3) is the common interface glued on the GB wall. The O-ring seat seals the glove port flange (2). The glove is secured to the flange through an extension spring blocking the forearm opening into the U-shaped slot surrounding the central hole. When not in use, the PMMA common interface is closed and sealed with the cover (1). All three elements are equipped with silicon rubber O-rings to maintain the seal of the assembly.}
\label{fig:figura_8}
\end{centering}
\end{figure}

\begin{figure}[h]
\begin{centering}
\includegraphics{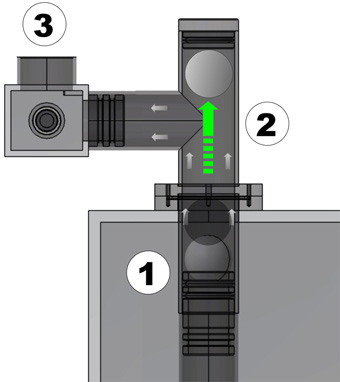}
\caption{Schematic view of the "ping pong" relief valve. In static conditions with a 4000 l/h flux, the ball floats halfway in the PMMA pipe (1). Overpressure peaks (the operator inserts his arm inside the GB) pull the ball to the top (2), fully opening the exhaust pipe. For a negative pressure, the ball is pushed on its seat, sealing the exhaust line. The picture shows an assembly with a zirconium oxide oxygen sensor (3) mounted at the valve outlet.}
\label{fig:figura_9}
\end{centering}
\end{figure}

\begin{figure}[h]
\begin{centering}
\includegraphics{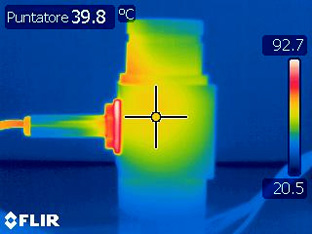}
\caption{Thermal image of the ZnO$_2$ Oxygen sensor mounted on its holder at the outlet of the relief valve.}
\label{fig:figura_10}
\end{centering}
\end{figure}

The assembly line design prioritizes the constraints posed by the recontamination issue and the need to match the CUORE requirements. However, it also advances a more general tool suitable for aiding the assembly of any other similar-sized structures requiring a strictly controlled clean environment and procedures. The switchable glove box design combined with the two-axis motion capability of the garage are an example of this effort; modular structures in a wide range of heights can be assembled inside the meck box, offering constant side and top views, while larger glove boxes, such as the cabling box, can present to the operator a fully extended, 1000-mm-tall assembly. Other elements of flexibility have been included in the glove box design using, whenever possible, common interfaces able to host different tools. One of the most effective is the circular access interface fixed at the production stage on the glove box wall. Their number and position vary with the glove box shape to provide an open standard sealed coupling system to any type of specialized flanges. Glove ports or electrical feed-through panels are examples of their use. The same glove boxes can be optimized for different assembly procedures by changing the type of the flanges installed on any available access.\\
The glove ports for CUORE have been designed to fix the gloves selected by the CUORE radioactivity group, the Trionic\textsuperscript{TM} 522, a triple polymer (a natural rubber neoprene and nitrile blend) gloves made by MAPA\textregistered Advantec. They are standard gloves available in two lengths, both sized for normal wearing (i.e., not shaped to be used as elements of a glove box system). Their size defines the working area accessible by the operators within a glove box and hence the UWP and the glove boxes' footprints. Figure ~\ref{fig:figura_8}
 shows the assembly of a long glove flange on the common PMMA interface. A slightly different glove flange is used to hold and fasten the short gloves. \\
The relatively small size of the glove boxes and the number of glove ports installed make the stability of the pressure inside the working volume strongly dependent on operators' activity. A differential pressure variation of a few millibars between the sealed volume and the outside environment could make it uncomfortable to wear gloves. Depending on the speed, an operator inserts or removes his arms from the box, and a variation of the sealed volume as fast as 2 l/s may occur. To ensure fast compensation of the negative pressure peaks, the glove boxes are fluxed up to 4000 l/h. A custom relief valve almost instantaneously levels the overpressure peaks and seals the exhaust pipe on the under-pressure peaks, thus keeping the box pressure almost constant at a few millibars above room level.\\
Commercial relief valves that are traceable on the market are not designed to operate concurrently at a few millibars overpressure and flow rates of several thousand liters per hour. The custom valve developed for our needs exploits the same operative principle found in many commercial products but uses as fast a pressure/flow regulator as a lightweight ping pong ball floating in a PMMA pipe. A standard ITT approved class ping pong ball has a 40.0$\pm$0.4 mm diameter, a sphericity greater than 0.25 mm and a mass of 2.72$\pm$0.03 g, which is enough to achieve the 5-6 mbar overpressure that we have tested to be the most comfortable for the operators. The pressure level inside the glove box can be further increased by adjusting the weight of the ball by filling it with some drops of Vaseline oil. \\
The relief valve (see figure ~\ref{fig:figura_9})
 fixtures allow the exhaust pipe line equipment to expand with different types of sensors to monitor the quality of the conditioned atmosphere. In the case of CUORE, the most important parameter to control is the $^{222}$Rn content.
The $^{222}$Rn activity measured in the clean room area is 25-30 Bq/m3, which implies the need of a suppression factor larger than 100 to match CUORE requirements. This level of activity cannot be monitored in real time. We chose to use oxygen measurements as an indirect, but fast, indication of the presence of air, and hence $^{222}$Rn, in the glove box volume.\\
Depending on the application, we use two types of oxygen sensors, a zirconium oxide one and a polarimetric one.\\
The zirconium oxide sensor is a cheap instrument with a sensitivity of 0.2\%  in volume, a fast response and negligible maintenance. Mounted at the relief valve outlet, it is used to monitor the purging cycle and the proper sealing of the glove box before opening the garage lid. Once reconditioning is over, the sensor is switched off since, by operative principles, it fails if exposed to a poor oxygen environment for a long time. From this moment on, the air contents in the garage glove box system are monitored by a higher sensitivity polarimetric sensor.\\
The zirconium sensor works at a high temperature (approximately 90$^{\circ}$C, see figure ~\ref{fig:figura_10})
 very close to the PMMA Vicat B softening temperature (100-110 $^{\circ}$C). Shielding, made by composite G-10, provides thermal decoupling at the PMMA holder. The sensor control board sends the data wirelessly to a simple wall-mounted receiver/display system. \\
The polarimetric sensor (Mettler Toledo InPro9050Gi) is a high-sensitivity O$_2$ sensor installed at the outlet of the garage relief valve. Its working principle, based on an O$_2$ induced chemical reaction, makes it the best choice to continuously monitor an almost O$_2$-free environment and gives a prompt response even for very small contamination. At the nominal rate of 800 l/h, the oxygen content inside the garage has been measured to be steadily less than 1 ppm.\\ 
The gas flow is controlled by four distribution boxes (the maximum number of gas independent volumes operated simultaneously during the assembly). The flow is set and monitored by a PLC controlled mass flow meter. The on-board software manages the safekeeping by blocking the flow in case of the wrong behavior of the system (for example, in the case of the glove box, the gas outlet or inlet is closed or blocked). 
\section{Mechanical assembly: the meck box and the EOW box }
\begin{figure}[h]
\begin{centering}
\includegraphics{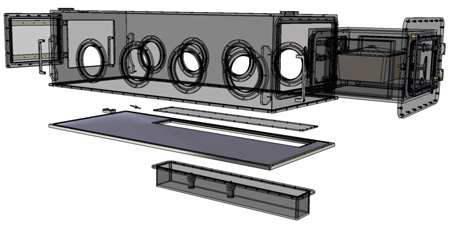}
\caption{Exploded view of the EOW box configured for parts selection and assembly feed. To speed up the parts throughput without spoiling the main volume N$_2$ atmosphere, a small antechamber replaces the right hand side closing panel. The antechamber is equipped with a set of empty, sealed, ancillary volumes to resize and keep to a minimum its effective volume to always get the faster reconditioning time.}
\label{fig:figura_11}
\end{centering}
\end{figure}

\begin{figure}[!]
\begin{centering}
\includegraphics{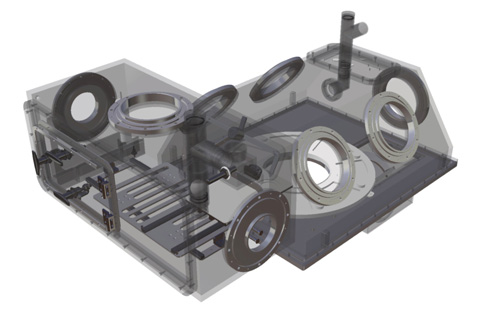}
\caption{The meck box. The antechamber (left) is glued to the glove box's main body. An interconnecting sealed door allows parts to pass from one volume to the other by means of a sliding tray. The antechamber body is raised with respect to the main chamber floor to allow access to the base flange fixtures and keep the sliding tray at a convenient height when entering the main volume. To always keep the inside of the Reber\textsuperscript{TM} box used to move the ultra-pure parts from the EOW box to the meck box clean, it is re-sealed by means of an external vacuum pump through a small pipe placed on the left wall of the antechamber.}
\label{fig:figura_12}
\end{centering}
\end{figure}

\begin{figure}[h]
\begin{centering}
\includegraphics{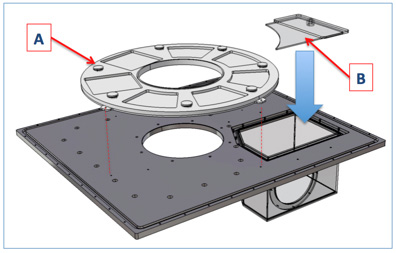}
\caption{The UWP with the PTFE rotating disk (A). Seven engraved slots hold ultra-clean parts. During mechanical assembly, a PMMA cover (B) closes access to the box under the UWP plane.}
\label{fig:figura_13}
\end{centering}
\end{figure}

\begin{figure}[h]
\begin{centering}
\includegraphics{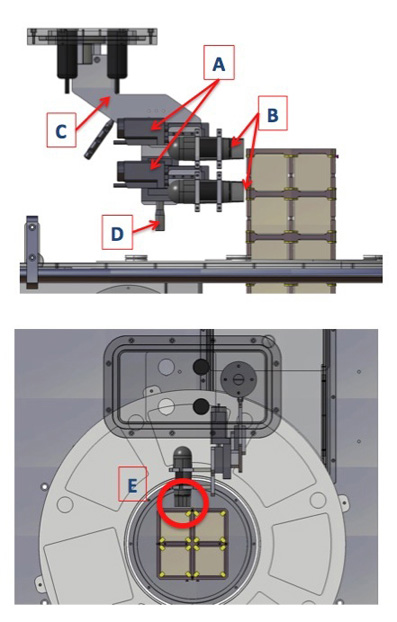}
\caption{The microscope systems. (A) The two step motors for focus adjustment; (B) the twin Dinolite\textsuperscript{TM} USB microscopes; (C) the rotating arm; (D) the micrometric screw; (E) top view of the microscopes pointing to the marks on two frames on a row.}
\label{fig:figure_14}
\end{centering}
\end{figure}

All of the ultra-pure detector components arrive packaged in clean containers and divided in small sets suitable to be used within the assembly line.\\
Crystals come packaged in groups of four inside sealed Reber\textsuperscript{TM} boxes under vacuum. PTFE spacers and copper parts packets are made of three polyethylene envelopes, each one individually set under vacuum. \\
Before entering the assembly process, the copper and PTFE parts are unpacked, individually inspected, tested and rearranged in smaller sets matching the assembly workflow. This task is performed inside the large volume, multi-access glove box (the "End of the Word" box in figure ~\ref{fig:figura_11}),
 whereas the meck box (figure ~\ref{fig:figura_12}) 
is the working volume where parts are assembled together. Both glove boxes are equipped with antechambers, which are small sealed pass-through boxes with a fast reconditioning time to provide a continuous stream of parts from the EOW box to the meck box. Both the inner and outer doors of the antechamber use quick Southco® fasteners for fast and easy opening. A natural rubber O-ring with the minimum shore value (35 Shore A Hardness) provides the proper sealing. \\
Ultra-pure component sets that pass quality controls are transferred "naked" (i.e., out of any envelopes) from the EOW box to the meck box antechamber inside clean PTFE canisters sealed in a dedicated vacuum-pumped Reber\textsuperscript{TM} box.\\
Once inside the meck box antechamber, the box is opened, and the canisters put on a clean sliding PTFE plate that, protruding from the meck box's main volume, makes them easily accessible to assembly operators. The clean components are then spilled inside one of the trays engraved on a large PTFE rotating disk surrounding the UWP central hole. Two operators access the components by spinning the disk by its handles (figure ~\ref{fig:figura_13}).\\
The assembly of each of the 13 floors of the tower is always performed at the same height, the most comfortable for the operator, thanks to the gradual lowering of the tower inside the garage. \\
The assembly of CUORE towers is a precision task. Pieces slightly out of spec or a loose coupling between two components may compromise the feasibility of later actions. In particular, a few-hundred-micron systematic deviation from the nominal distance among frames may prevent the correct coupling between the tower and the wire trays carrying the PCB cabling. \\
A method to monitor the structure growth without spoiling the radio purity has been developed by exploiting the garage's vertical motion and designing an optical system capable of measuring the distance between two frames without touching them. The device is made by two superimposed Dinolite\textsuperscript{TM} USB microscopes (magnification up to 200x) fixed to a rotating arm hanging from the top of the meck box (figure ~\ref{fig:figure_14}).
The distance between the two microscopes is fine adjusted by means of a micrometric screw and a calibrated target, while focusing is individually set by operating two micro step motors (1,5 $\mu$m/step) that drive their forward and backward displacement. The detector growth is then monitored floor by floor by measuring the distance between two frames and comparing it to the nominal value. By rotating the tower, the same measure can be repeated on the four sides of the assembly. At a magnification of 200x, the sensitivity on the distance between two marks is approximately 10 $\mu$m.\\
The system is fixed to a special square flange, placed on top of the meck box's main body, and carries the USB and the step motor signals feed-through plugs.\\
Once the mechanical assembly is over, the tower is placed back into the garage, and the meck box is removed to prepare the UWP. 
\section{Cabling: the cabling box and the EOW box }
\begin{figure}[!]
\begin{centering}
\includegraphics{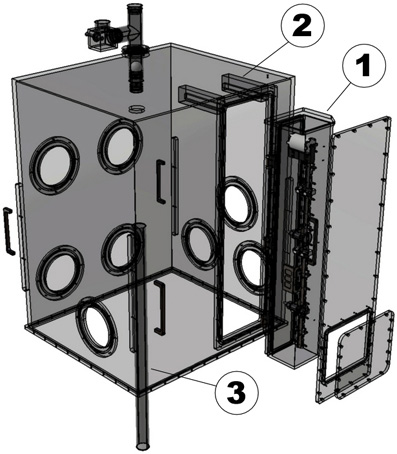}
\caption{Exploded view of the cabling box. The picture shows the transfer box (1) ready to enter the main volume. The top of the transfer box fits a slide (2) on the cabling box top, which ensures the stable vertical position of the tool. Handles on the four corners of the box make it easy to put on or remove it from the UWP. The gas flows from the top of the box from an auto-sealing plug and is drained out from the bottom through a pipe (3) connected to the relief valve.}
\label{fig:figura_15}
\end{centering}
\end{figure}

\begin{figure}[!]
\begin{centering}
\includegraphics{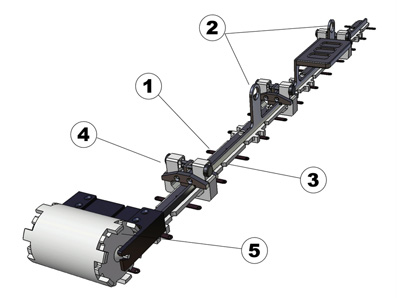}
\caption{The exoskeleton is fastened to the wire tray (1) by means of three PTFE clamps (4). Two wings (2) provide the interfaces to the transfer box holder and the tower alignment tool. The white clean PTFE slipper (3) avoids contact between the clean components and the aluminum stand. The flexible PCB cabling left over for the detector connection to the front end electronics is folded inside a clean PTFE roll (5) secured on the top of the exoskeleton.}
\label{fig:figura_16}
\end{centering}
\end{figure}

\begin{figure}[h]
\begin{centering}
\includegraphics{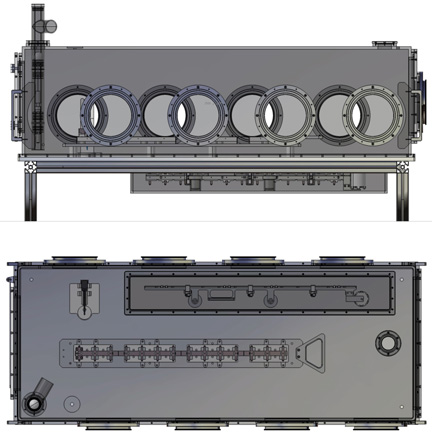}
\caption{: Top and side view of the EOW box configured for wire tray assembly.
The antechamber is removed to maximize the working area.
The asymmetric placement of the glove access on the two sides of the box eliminates blind zones on the working plane. The side view shows the cellar hanging from the working plane's bottom and hosting the exoskeleton while waiting for the wire tray assembly to be completed. It can be a sealed and stored in a safe location during an emergency while having clean assembly.}
\label{fig:figura_17}
\end{centering}
\end{figure}

The "cabling box" (figure ~\ref{fig:figura_15})
size and access layout allow for the hosting and operation of a fully exposed 934-mm-tall CUORE tower structure. The main side is equipped with two couples of glove ports placed at two different heights. The left and right sides are equipped, respectively, with one and three interface flanges placed asymmetrically. They can be set as glove accesses or hold cable pass-through ports depending on the assembly action to be performed. On the backside, a tall door allows for the insertion of a special sealed box (the "transfer box") sized and equipped to hold pre-assembled parts up to 1000 mm long. A smaller door, engraved into the taller one, is used to let in smaller loads.\\
In the CUORE tower assembly process, the transfer box is used to couple the tower with the thin (1 mm thick) and long (934~mm) copper piece (wire tray) holding the flexible PCB strips carrying out the bolometer signals.\\
The wire tray-PCB coupling is performed within the EOW box. The antechamber is removed and the work plane is equipped with the clean tools needed to align and glue the ten layer PCB strip pack to the four copper fishbone-shaped sections that constitute the wire try structure. This delicate assembly is then secured to a clean and stiffer bar called the "exoskeleton" (Figure ~\ref{fig:figura_16}).
 The exoskeleton allows the assembly to rotate vertically and provides the interface to the tools used to align and couple the wire tray to the tower.\\
While waiting for the wire tray assembly to be completed, the exoskeleton is stored inside an auxiliary sealed volume (the "cellar"), extending beneath the EOW work plane and accessible from the inside only (Figure ~\ref{fig:figura_17}).
 Once the glue is cured, the exoskeleton is pulled out, coupled with the wire tray by means of a sliding clamps system, put back and sealed into the cellar again. This sealed temporary storage protects the clean assembled part while the EOW box is opened and cleared to insert the transfer box used to move the wire trays from the EOW box to the cabling box.\\
Once inside the cabling box, the canister remains fastened until the main volume purging cycle ends. After purging, the transfer box is opened, and the exoskeleton is locked to a revolving stand that helps the operator move toward and align the wire trays to the tower. Once the wire tray is fastened to the tower, the exoskeleton is disengaged, making the detector ready for bonding. 
\section{The Bonding Box}
\begin{figure}[!]
\begin{centering}
\includegraphics{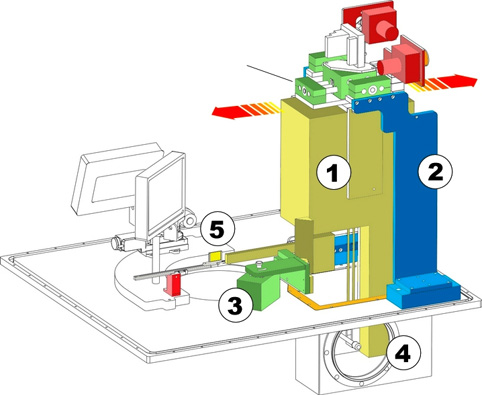}
\caption{: Front view of the UWP set for bonding: (1) the wire vertical bonder; (2) the support structure coupled with the machine top; (3) the LCD video camera joint with the wire bonder's body; (4) the access to the manual drive beneath the UWP; (5) a gold target holder sliding in front of the machine for parameter tuning and testing.}
\label{fig:figura_18}
\end{centering}
\end{figure}

\begin{figure}[!]
\begin{centering}
\includegraphics{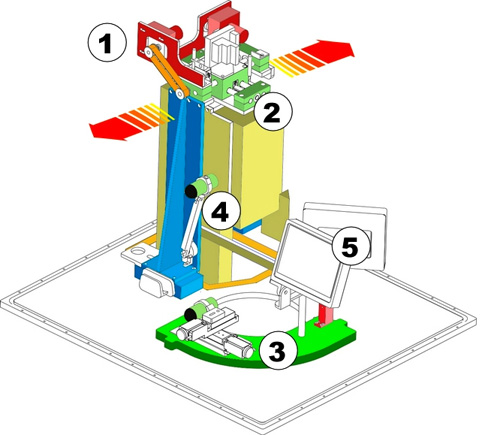}
\caption{: Rear view of the UWP set for bonding: (1) one of the two DC motors driving the bonder motion. The motor axis is connected to the translation stage by a cog belt for a smoother drive (2); the wire bonder x,y translation stage; (3) the half-moon plate carrying the rear QC microscope; and (5) the two LCD monitors on their adjustable stand. (4) A magnetic clamp easily fastens the microscope to the x,y positioning table or to a rotating arm (4), allowing the bonder tip to be framed for maintenance actions.}
\label{fig:figura_19}
\end{centering}
\end{figure}

\begin{figure}[h]
\begin{centering}
\includegraphics{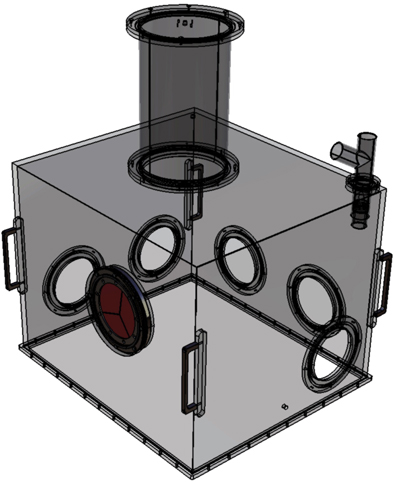}
\caption{The bonding box.} 
\label{fig:figura_20}
\end{centering}
\end{figure}

The UWP is designed to host a 7700 West Bond manual wire bonder modified by the manufacturer to operate on a vertical plane to make feasible the bonding of the components lying on the tower surface.\\
The standard area covered with the manual drive of the bonding tip ($\pm$10 mm left-right) has been extended by hanging the top of the machine to a 70-mm, two-axis, motorized table fixed to a bridge-shaped aluminum fixture just above the shaped opening of the UWP. This system, along with the z axis garage motion, reaches any bonding target within $\pm$45 mm (left-right) and $\pm$10 mm in depth from the detector center all along the 934-mm-tall tower. Two DC motors controlled by proportional actuators provide a smooth and precise drive of the instrument (figure ~\ref{fig:figura_18}).
  \\
The moveable structure integrates an LCD camera that continuously frames the area beyond the tip. Focus and magnification remote controls help the operator aim and reach with high precision the bonding target.\\
Opposite to the bonding system (figure ~\ref{fig:figura_19}),
  the UWP is equipped with an interface (the half-moon plate) hosting a two axis motorized stage that holds a microscope for bonding target inspections and bonding quality check plus two small steerable LCD monitors serving the microscope and the main LCD camera.
All of these instruments are included in the bonding box shown in figure \ref{fig:figura_20}.
   \\
The glove box provides two glove accesses on the rear side (cabling, tools installation, and quality check operations) and a cable pass-through flange on the right side. The onboard machine's front panel manual switches can be reached both from a fast access IRIS port on the front side of the box and from a glove access on the left side, which is suitable for machine maintenance actions, too.\\
All of the cable pass-through connections (the main LCD camera video-out, the bonding machine power, the x/y wire bonder motion, and the x/y motorized control of the rear QC microscope) are IP68/69K industrial standard compliant, which ensures, when mated, proper sealing of the glove box.
The shape of the box is designed to minimize its volume by limiting to a cylindrical extension the volume exceeding the minimum set by the wire bonder heights and its equipment.\\
The bonding machine is a complex electro-mechanical instrument that is hard to keep as clean as all of the other tools installed inside a glove box. Even if stored in a clean room, dirt and contaminants may collect in the interior whenever it is out of the glove box. To avoid that, the gas flow crossing this area goes towards the detector, and the N$_2$ inlet is placed on top of the cylindrical extension, steering the flow straight over the tower while the exhaust pipe draws gas from the glove box's bottom, near the corner behind the machine itself.\\
A shower-like diffuser at the inlet slows down the flow to minimize turbulence that may interfere with the bonding operation.
\section{Removing and Storing: the Storage Box}
\begin{figure}[h]
\begin{centering}
\includegraphics{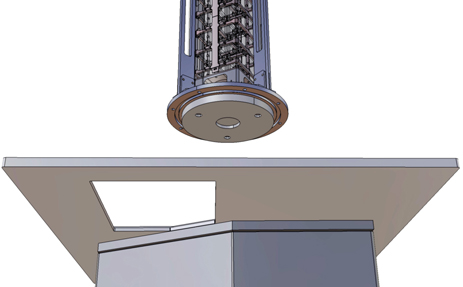}
\caption{: The tower leaving the garage inside the storage box. The same plate the detector has been assembled on closes the bottom of the canister.}
\label{fig:figura_21}
\end{centering}
\end{figure}

\begin{figure}[h]
\begin{centering}
\includegraphics{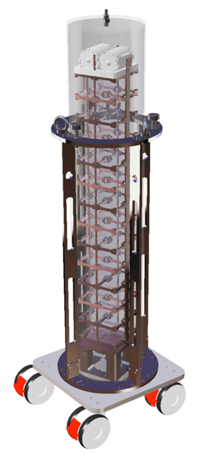}
\caption{: The storage box on its cart.}
\label{fig:figura_22}
\end{centering}
\end{figure}

\begin{figure}[h]
\begin{centering}
\includegraphics{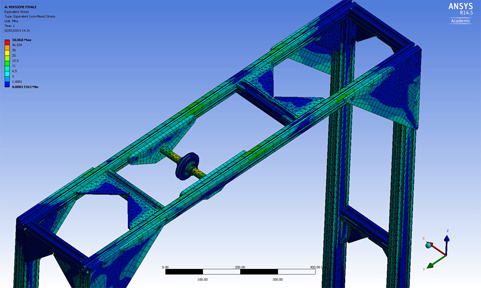}
\caption{: Ansys calculation of the stress and strain of the crane.}
\label{fig:figura_23}
\end{centering}
\end{figure}

Once the assembly is completed, the detector is removed from the garage and is inserted into a sealed canister (the storage box).\\
This action breaks the zero-contact approach pursued throughout the previous assembly stages, exposing the tower, even for a very short time, to an atmosphere that is not completely purged. To limit risks when the garage lid is open and to allow the detector to move inside the canister, the storage box is designed as a volume open at the bottom, using the tower's assembly plate itself as the sealing shutter (figure~\ref{fig:figura_21}). 
 \\
The canister base flange has a double circular hole pattern, the outer to be coupled with the UWP and the inner to fasten the assembly plate. \\
Two NBR, 90 Shore A hardness, O-rings, one on the base flange bottom and the second one on the assembly plate's top, give the two volumes (the canister and the garage) tightness while relocating the detector and guarantee the air tightness of the coupling of the assembly plate with the storage box base flange.\\
To prepare and keep the storage box volume clean, both before the lid opens and during the transfer, it is continuously flushed with nitrogen at a rate of 4000 l/h. During this action, the risk to air exposure is related to the garage lid opening and the resulting contamination of its atmosphere. The polarimetric sensor reading shows that the oxygen contents (air =100\%) inside the garage reach, at maximum, the 13\% level, lasting over the 1\% level for less than 3 minutes, whereas the detector move takes about one minute.\\
The box that stores the CUORE tower is an aluminum alloy structure (figure~\ref{fig:figura_22})
 with a PMMA cylindrical volume equipped with gas inlet-outlet plugs. Four light standards offer breaks-in to pick up the canister with different tools, while transparent housing allows for the visual inspection of the tower from a very close point of view. The bottom flange fits a specific wheel cart used to propel the box around the clean room and to maintain the cleanness of the assembly plate's bottom.\\
The mass of the storage box with the tower is approximately 60 kg; in order to provide a smooth removal from the UWP, a light handheld portal crane with a manual winch has been designed for the job. It is a modular structure made of aluminum extrusion bars (FBV Modular System products, series 30x30 mm$^2$ and 30x60 mm$^2$ ) calculated and tested to lift up to 85 kg. A FEM static-structural analysis in ANSYS (figure~\ref{fig:figura_23})
  \\
 shows the maximum stress in the aluminum extrusion as 58 MPa, calculated following the Von Mises criterion; it achieves the SF=1.5 considered for a dynamic load compared to the yield stress certified for the extrusion material (190 MPa). When not in use, its compact shape and foldable legs make it easy to park it in the clean room without spoiling the little space available. 
\section{Conclusions}

In this paper, we describe the construction strategy and the procedures conceived to assemble the 19 towers that constitute the CUORE detector.
The CTAL in the CUORE clean room at LNGS was commissioned in the summer of 2011.\\
A first CUORE-like tower, CUORE-0, was assembled in March 2012 as a bench test to validate the CUORE production chain sharing the same raw material selection guidelines, parts machining protocols, cleaning techniques and assembly procedures of a standard CUORE tower. The Cuore-0 construction was terminated in 5 working days (wire bonding not included).\\
During this test, the intrinsic flexibility of the system was widely tested and proved to be very effective in making this delicate and ultra clean assembly feasible, fast and easy.\\
The complete reversibility of the process also makes it possible to perform many quality checks on the parts and their coupling prior to the final assembly and without spoiling their radio purity.\\
CUORE-0 is currently operated at ~13 mK inside the Cuoricino cryostat in the LNGS hall A. A first look ~\cite{8} at the data taken in the very first few months shows a significant improvement on the background suppression in the ROI with respect to the Cuoricino results, confirming that the performance of this system is compliant with CUORE requirements.\\
With a few minor modifications and upgrades, as suggested by the Cuore-0 experience, CTAL started to assemble the CUORE towers in March 2013. The construction has proceeded smoothly since then, in line with the CUORE construction master plan.\\
At the time of writing, 18 of the 19 towers have been assembled without problems. Construction is expected to end in June 2014.



\bibliographystyle{elsarticle-num}
\bibliography{<your-bib-database>}



\end{document}